\documentclass{article}

\usepackage{arxiv}

\usepackage[utf8]{inputenc} 
\usepackage[T1]{fontenc}    
\usepackage{hyperref}       
\usepackage{url}            
\usepackage{booktabs}       
\usepackage{amsfonts}       
\usepackage{nicefrac}       
\usepackage{microtype}      
\usepackage{lipsum}		
\usepackage{graphicx}
\usepackage{doi}

\usepackage{caption}
\usepackage{subcaption}

\usepackage{framed,multirow}
\usepackage{amssymb}
\usepackage{latexsym}
\usepackage{commath}
\usepackage{amsmath}
\usepackage{lscape}
\usepackage{siunitx}
\usepackage{appendix}
\usepackage{mathabx}

\title{Block Spectral Stresses (BSS) estimation for shock-capturing and turbulent modeling}

\author{{Matteo Ruggeri}\\
	Department of Aeronautics \& Astronautics Engineering\\
	Purdue University\\
	\texttt{mruggeri@purdue.edu} \\
	\And
	{Victor C. B. Sousa}\\
	Department of Mechanical Engineering\\
	Purdue University\\
	\And{Carlo Scalo}\thanks{Department of Aeronautics \& Astronautics Engineering by courtesy}\\
	Department of Mechanical Engineering\\
	Purdue University\\
}

\renewcommand{\shorttitle}{\textit{arXiv} Template}

\hypersetup{
pdftitle={A template for the arxiv style},
pdfsubject={q-bio.NC, q-bio.QM},
pdfauthor={David S.~Hippocampus, Elias D.~Striatum},
pdfkeywords={First keyword, Second keyword, More},
}

\begin{document}
\maketitle

\begin{abstract}
A new combined sub-filter scale turbulence and shock-capturing model is developed for high-order finite volume numerics, extending previous work \cite{LSV} to unstructured solvers. Block Spectral Stresses (BSS) method relies on the spectra of the velocity gradients to estimate the subfilter scale stresses, heat-flux, and pressure-work based on the resolved field. The method is able to capture shocks with numerical order up to 25 and in a shock-vortex interaction simulation is able to capture the shock and not interfere with the vortex structure. In turbulence calculations the new method is compared with Smagorinsky, dynamic Smagorinsky, and Vreman methods adapted to a block spectral code. In the simulations of homogeneous isotropic turbulence, the new model is worse than the others when on coarse meshes and better on finer ones. Instead, for supersonic and hypersonic channel flow the case is the opposite because as expected the sub-filter terms are mostly depend on the numerical order and not the mesh resolution.

\end{abstract}


\section{Introduction}
\label{sec:intro}

The last years have seen the proliferation of pseudospectral methods which merge the benefits of unstructured mesh to the high-order numerics which allow a better resolution than low order methods with a smaller number of degrees of freedom, given their faster convergence rate \cite{godunov1959finite,toro2013riemann,FR,hunyh2,hunyh3}. Indeed, pseudospectral methods show a convergence rate closer to spectral methods than finite difference ones. But at the same time they shows the same limitation \cite{gibbs1,gibbs2,gottlieb1977numerical} of the high-order method when it has to work with steep gradients as in the case of shocks in compressible flow. In addition to shocks in gas dynamic flow another phenomenon that can cause numerical instability and leads to the blow-up of the simulation if not properly modeled is turbulence. Both turbulence and shock formation have a similar energy cascade from large to small scales because of nonlinear interaction \cite{QSV,gupta_lodato_scalo_2017}. Differently from a shock that has an infinite spectra, turbulence has a finite spectra therefore standard shock-capturing methods usually tend to remove a part of the spectra of the turbulence and therefore reduce the quality of the simulation. Instead, Eddy viscosity models--which allow to resolve turbulence on a coarser mesh than Direct Numerical Simulation quality (DNS) thanks to the modeling of small scales turbulence--are not able to fully dissipate the shock. This work aims to address these two problems with a unified model that is able to capture the shock and the same time does not remove a broad spectra of the turbulence from the simulation.

Block spectral methods can be implemented with three different approaches: the discontinuous Galerkin (DG) \cite{BASSI1997267,DG}, the staggered grid approach (or spectral differences) \cite{KOPRIVA1996244}, and the flux reconstruction (FR) method \cite{FR}. The flux reconstruction (FR) method is the latest to be developed and it relies on correction functions that are used to interpolate back to the solution points the fluxes from their updated value at the interfaces, this makes them continuous among elements. Hyunh \cite{FR} pointed out that the flux reconstruction method is depend to the functions that are use for correction and hence it can be made more or less smooth, stable, or accurate than other numerical methods. Asthana et al. \cite{ASTHANA2015269} proved that even when the computational domain is refined the instabilities caused by aliasing errors related to the collocation projection of the flux at the jump conditions are not completely reduced. In the same paper, they mathematically proved that if an artificial viscosity is added to flux reconstruction method, it can be  stabilized with all numerical order after a certain grid refinement. They also showed how this allows to run 1D shock simulation up to numerical order 120 and even without the sensor that detects the shock the rate of convergence is not much different from when is active (the same order of accuracy was achieved by Sousa and Scalo \cite{LSV} for 1D Sod shock tube).

Tonnicello et al. \cite{TONICELLO2020104357} addressed the problem of preserving entropy in a shock when applying the shock-capturing method in a high-order spectral difference numerics code. The entropy profile across a shock is not monotonic (also shown by Gupta and Scalo \cite{gupta2018spectral}) but it presents a peak inside the propagation wave this result has been shown by Morduchow and Libby \cite{morduchow} and later validated by other works \cite{smoller2012shock,salas1996entropy,colombeau1988multiplications}. The problem is that most of the shock-capturing methods are not able to capture this phenomenon and make the entropy monotonic across the shock. They compare the classical Laplacian viscosity to a physical artificial viscosity method, showing that the latter is able to have a better prediction of the entropy even when the Laplacian approach is able to resolve the shock well. The main drawback of this approach is that it also removes the spectra of the acoustic waves, which in the case of supersonic and moreover in hypersonic flow have great importance because phenomena like second mode waves lead the transition to turbulence in hypersonic flow. The authors pointed out that the suppression of acoustic waves can be caused by the discontinuity-based sensor use to the detect and shock and added that a divergence-based sensor may have better performance. Nevertheless, the authors presented works \cite{fernandez2018physics,ducros1995large} that use a divergence-based sensor and they still have the removal of acoustic waves when compare to DNS simulations.

The current work aims to address this challenge in developing a combined SFS turbulence shock-capturing model that is able to capture the shock, while retaining the spectrum of turbulence. Section \ref{sec:model_desc} describes the code used for this investigation and the numerics used to solve the compressible Navier-Stokes equations. In the second part of the section the Block Spectral Stresses (BSS) model is introduced together with a description of the Smagorinsky and dynamic Smagorinsky models implemented in a block spectral code and later use for comparison with BSS. Section \ref{sec:BSS_shock} presents how the model performs in 1D and 2D flow with shocks. Then the model is tested to act as an LES model on a Taylor-Green Vortex (TGV) flow and turbulent wall-bounded flow at supersonic and hypersonic flow speed in section \ref{sec:LES_comp}.

\section{Models description}
\label{sec:model_desc}

The code used to perform the simulations is H$^3$AMR \cite{AFRLjournal,porous_scitech,porous_aviation,2023_RunningEtAl_EXiF} (HySonic, High-Order, Hybrid Adaptive Mesh Refinement developed by HySonic Technology, LLC) an unstructured block-spectral code \cite{H3AMRnum} for compressible flow based on flux reconstruction \cite{FR} to compute the gradients. In the current section a description of the code \ref{subsec:hammer} and of the LES models as implemented in the code are provided: Smagorinsky \cite{smago} (Section \ref{subsubsec:sma}), Dynamic Smagorinsky \cite{dynamicSma} (Section \ref{subsubsec:dyn}), Vreman's model \cite{vreman} (Section \ref{subsubsec:vreman}), and Block Spectral Stresses (BSS) (Section \ref{subsubsec:BSS}).

\subsection{Code description (H3AMR)}
\label{subsec:hammer}

The vector of conserved quantities reads:

\begin{equation}
    \mathbf{Q} = \begin{bmatrix}
           \rho \\
           \rho u_1 \\
           \rho u_2 \\
           \rho u_3 \\
           \rho E
         \end{bmatrix}
    \label{eq:Q_cons}
\end{equation}

\noindent where $\rho$ is the density, $u_i$ the velocity in the $i^{th}$ direction, and $E=e+\frac{u_i u_i}{2}$ is the specific total energy ($e=\frac{p}{\rho (\gamma-1)}$ is the specific internal energy in the case of ideal calorically perfect gas). The flux vectors read:

\begin{equation}
\begin{aligned}
   & \mathbf{F}= & & \mathbf{G}= & & \mathbf{H}= & \\
   & \begin{bmatrix}
           \rho u_1  \\
           \rho u_1 u_1 + p - \mu \sigma_{11} \\
           \rho u_1 u_2 - \mu \sigma_{12} \\
           \rho u_1 u_3 - \mu \sigma_{13} \\
           \rho (\rho e + p)u_1 - k \frac{\partial T}{\partial x_1} + \\ \dots- \mu \sigma_{i1} u_i
         \end{bmatrix} & & \begin{bmatrix}
       \rho u_2  \\
       \rho u_2 u_1 - \mu \sigma_{21} \\
       \rho u_2 u_2 + p - \mu \sigma_{22} \\
       \rho u_2 u_3 - \mu \sigma_{23} \\
       \rho (\rho e + p)u_2 - k \frac{\partial T}{\partial x_2} + \\ \dots - \mu \sigma_{i2} u_i
     \end{bmatrix} & & \begin{bmatrix}
       \rho u_3  \\
       \rho u_3 u_1 - \mu \sigma_{31} \\
       \rho u_3 u_2 - \mu \sigma_{32} \\
       \rho u_3 u_3 + p - \mu \sigma_{33} \\
       \rho (\rho e + p)u_3 - k \frac{\partial T}{\partial x_3} + \\ \dots - \mu \sigma_{i3} u_i
     \end{bmatrix} & \\
     \label{eq:FGH}
\end{aligned}
\end{equation}

\noindent where $\mu$ is the dynamic viscosity, $p$ the pressure, $T$ the temperature, and $\sigma_{ij}=\frac{\partial u_i}{\partial x_j}$ the strain-rate tensor. Given these vectors, the Navier-Stokes equations can be written as:

\begin{equation}
    \frac{\partial \mathbf{Q}}{\partial t} + \frac{\partial \mathbf{F}}{\partial x_1} + \frac{\partial \mathbf{G}}{\partial x_2} + \frac{\partial \mathbf{H}}{\partial x_3} = 0
    \label{eq:NS_phy}
\end{equation}

\noindent and they are still in the physical space. Considering that the numeric is specifically designed for a computational space the conserved quantities vector and the flux vectors can be recast in computational space, with coordinates $\xi_i$ through:

\begin{equation}
\begin{aligned}
    Q =  J\mathbf{Q} \\
    F =  J\left(\frac{\partial \xi_1}{\partial x_1}\mathrm{F} + \frac{\partial \xi_1}{\partial x_2}\mathrm{G} + \frac{\partial \xi_1}{\partial x_3}\mathrm{H} \right) \\
    G =  J\left(\frac{\partial \xi_2}{\partial x_1}\mathrm{F} + \frac{\partial \xi_2}{\partial x_2}\mathrm{G} + \frac{\partial \xi_2}{\partial x_3}\mathrm{H} \right) \\
    H =  J\left(\frac{\partial \xi_3}{\partial x_1}\mathrm{F} + \frac{\partial \xi_3}{\partial x_2}\mathrm{G} + \frac{\partial \xi_3}{\partial x_3}\mathrm{H} \right) \\
\end{aligned}
     \label{eq:transform}
\end{equation}

\noindent where the italics characters are considered in computational space, $\frac{\partial \xi_i}{\partial x_j}$ is the linear transformation matrix from the physical $x_j$ to computational space $\xi_i$ and $J$ the Jacobian of the transformation matrix. Therefore the Navier-Stokes equations can be rewritten as:

\begin{equation}
    \frac{\partial \mathbf{Q}}{\partial t} + \frac{1}{J}\frac{\partial F}{\partial \xi_1} + \frac{1}{J}\frac{\partial G}{\partial \xi_2} + \frac{1}{J}\frac{\partial H}{\partial \xi_3} = 0
    \label{eq:NS_comp}
\end{equation}

\noindent which are the equations solved by H3AMR.

\begin{figure}[h!]
    \centering
\includegraphics[width=0.8\textwidth]{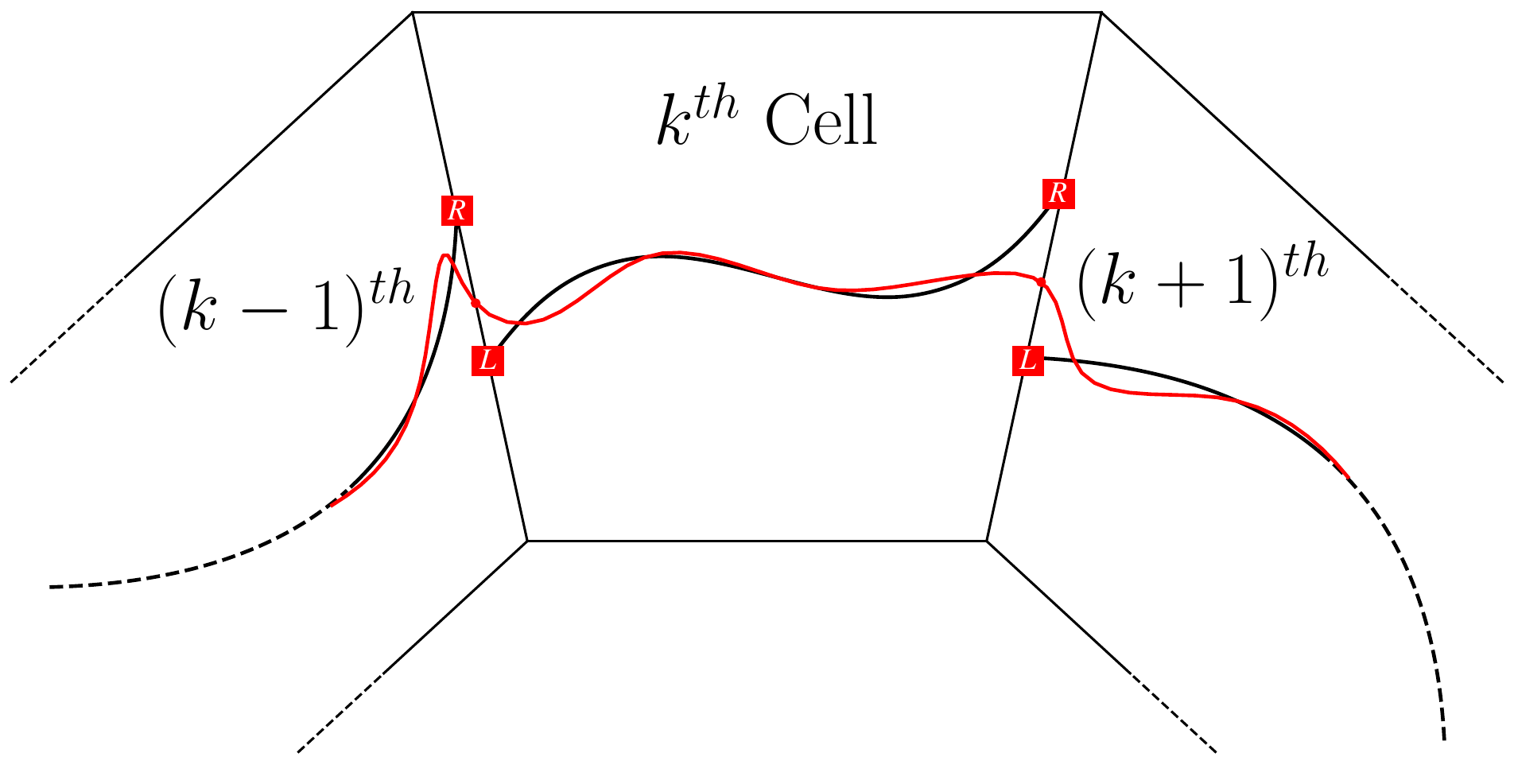}
    \caption{Example of 1D elements which are interpolate to the interface (black lines) and then corrected back to the solution points with their updated values at the interface (red line).}
    \label{fig:FR}
\end{figure}

The computational domain is a mesh divided into several elements, in each element, the conserved quantities and the fluxes are stored in the solution points which are the Gauss-Legendre quadrature points (N+1 points given a polynomial reconstruction of order N). From now on every reference to numerical order is equal to the number of solution points in each cell and hence the polynomial order is the numerical order minus one. For the simplicity of exposure, all the following numerics will be described in a 1D formulation, but it is enough to use an outer product to extend it to a 3D domain. Given a set of values of conserved quantities and fluxes in the solution points $f(\xi_j)$ their values are interpolated to the interfaces of the element ($f(-1)$ for the left interface and $f(+1)$ for the right interface) and then updated to their new value using the Rusanov \cite{Rusanov} method:

\begin{equation}
    \mathrm{F}^{new} = \frac{1}{2}(\mathrm{F}_R + \mathrm{F}_L) - \frac{S^+}{2}(\mathrm{Q}_R - \mathrm{Q}_L)
    \label{eq:rusanov}
\end{equation}

\noindent where $F^{new}$ is the new value of the flux at the interface, $F_R$ and $F_L$ are the initial values of the fluxes at the right and left of the interface, $Q_R$ and $Q_L$ are the initial values  of the conserved quantities at the right and left of the interface, and $S^+$ is the maximum between the sum of the speed of sound $a$ and velocity $u$ at the left and right of the interface. From the updated values the derivatives can be computed using the flux reconstruction \cite{FR} method:

\begin{equation}
    \frac{\partial f(\xi_j)}{\partial \xi}= D_{ij}f(\xi_j) + \left(f^{un}_{-\frac{1}{2}}-f(-1) \right)g'_L(\xi_i) + \left(f^{un}_{+\frac{1}{2}}-f(+1) \right)g'_R(\xi_i)
    \label{eq:dev_FR}
\end{equation}

\noindent where $f^{un}_{-\frac{1}{2}}$ and $f^{un}_{+\frac{1}{2}}$ are the updated values at the left and right interface of the element, $D_{ij}$ is the matrix representation of the first-order derivative, and $g'_L(\xi_i)$ and $g'_R(\xi_i)$ are the derivatives of the correction functions on the solution points. The correction functions have the property of being symmetric $g_R(\xi_j)=g_L(-\xi_j)$ and they can be computed from the Radau polynomials:

\begin{equation}
    R_{R,N+1}(\xi_j) = \frac{(-1)^{N+1}}{2}\left(L_{N+1}(\xi_j) - L_N(\xi_j)\right),
    \label{eq:radau}
\end{equation}

\noindent where $R_{R,N+1}(\xi_j)$ and $L_{N+1}(\xi_j)$ are the Radau and Legendre polynomials of polynomial order $N+1$ at $\xi_j$. From the Radau polynomials, the correction functions can be computed as:

\begin{equation}
    g_{N+1} = \frac{N+1}{2N + 1} R_{R,N+1} + \frac{N}{2N + 1} R_{R,N}.
    \label{eq:corr_funs}
\end{equation}

Figure \ref{fig:FR} shows how this process works, where the black line shows the interpolation to the interface of the element. Then, the values at the interface are updated to a new common value which is the red dot at the interface. The updated quantity is interpolated back to the solution points making the reconstruction inside the element $C^0$ continuous among elements.

\subsection{Filtered Navier-Stokes formalism}
\label{subsec:LES}

The Navier-Stokes equations in physical space can be filtered with a spatial filter that can commute with the derivative operation, a variable with the bar on top is considered filtered $\bar{f}$. For compressible flows it is suggested to use a Favre-based filter: 

\begin{equation}
    \check{f}=\frac{\overline{\rho f}}{\bar{\rho}}, 
    \label{eq:favre_filt}
\end{equation}

\noindent which leads to the Favre-filtered Navier-Stokes equations:

\begin{equation}
    \frac{\partial \overline \rho}{\partial t} + \frac{\partial \overline {\rho} \check{u}_j}{\partial x_j} = 0
    \label{eq:NS_LES_cont}
\end{equation}

\begin{equation}
    \frac{\partial \overline{\rho}  \check{u}_i}{\partial t} + \frac{\partial}{\partial x_j} \left( \overline \rho \check{u}_i \check{u}_j  +  \overline{p} \delta_{ij} -  \mu \check{\sigma}_{ij} +  \overline \rho \tau_{ij} \right) = 0,
    \label{eq:NS_LES_mom}
\end{equation}

\begin{equation}
    \frac{\partial \overline E}{\partial t} + \frac{\partial}{\partial x_j} \left( (\overline E + \overline p) \check u_j - k \frac{\partial \check T}{\partial x_j}  - \mu \check \sigma_{ij} \check u_i +  \frac{1}{2} \left( \frac{\gamma \pi_j}{\gamma - 1}  + \overline \rho C_p q_{j} \right) +  \frac{1}{2} \overline \rho \zeta_{j}  \right) = \mu \epsilon,
    \label{eq:NS_LES_energy}
\end{equation}

\begin{equation}
    \frac{\overline p} {\gamma - 1} = \overline E -  \frac{1}{2} \overline \rho \check u_i \check u_i - \frac{1}{2} \overline \rho \tau_{ii}.
    \label{eq:ideal}
\end{equation}

\noindent The last is the state equation in the case of an ideal gas, $k$ and $C_p$ are the thermal conductivity and heat capacity at a constant pressure of the gas. Among the nonlinear terms ($\tau_{ij}$, $q_j$, $\pi_j$, $\zeta_j$, and $\epsilon$) the SFS kinetic energy advection $\zeta_j$ and the SFS heat dissipation $\epsilon$ can be considered negligible. The Smagorinsky and Dynamic Smagorisky models assume negligible also the SFS pressure-work $\pi_j$, which is instead modeled by the BSS method because leads to better results when the model is used for shock-capturing. The quantities can be analytically computed as:

\begin{equation}
    \tau_{ij} = \widecheck{u_i u_j} - \check u_i \check u_j
    \label{eq:tau_SFS}
\end{equation}

\noindent for the SFS stress tensor, 

\begin{equation}
    q_{j} = \widecheck{T u_j} - \check T \check u_j
    \label{eq:q_SFS}
\end{equation}

\noindent for the SFS heat flux, and

\begin{equation}
    \pi_{j} = \overline{p u_j} - \overline p \check u_j
    \label{eq:pi_SFS}
\end{equation}

\noindent for the SFS pressure-work. Considering that BSS is a purely numerical method all the gradients are computed in computational space and to have a similar comparison also the other models use the same gradients. The Smagorinsky models are adapted to the computational gradients and considering that the computational domain is perfectly orthogonal in this specific case the predicted SFS quantities would be the same in physical and computational space. The SFS quantities are then added to the physical flux vectors:

\begin{equation}
    \mathbf{F}^{\mathrm{SFS}}_j = \begin{bmatrix}
           0  \\
           \bar{\rho} \tau_{11} \\
           \bar{\rho} \tau_{12} \\
           \bar{\rho} \tau_{13} \\
           \frac{1}{2} \left( \frac{\gamma \pi_1}{\gamma - 1}  + \overline \rho C_p q_{1} \right) 
         \end{bmatrix}
    \mathbf{G}^{\mathrm{SFS}}_j = \begin{bmatrix}
       0  \\
           \bar{\rho} \tau_{21} \\
           \bar{\rho} \tau_{22} \\
           \bar{\rho} \tau_{23} \\
           \frac{1}{2} \left( \frac{\gamma \pi_2}{\gamma - 1}  + \overline \rho C_p q_{2} \right) 
     \end{bmatrix}
    \mathbf{H}^{\mathrm{SFS}}_j = \begin{bmatrix}
       0  \\
           \bar{\rho} \tau_{31} \\
           \bar{\rho} \tau_{32} \\
           \bar{\rho} \tau_{33} \\
           \frac{1}{2} \left( \frac{\gamma \pi_3}{\gamma - 1}  + \overline \rho C_p q_{3} \right) 
     \end{bmatrix}
     \label{eq:FGH_BSS}
\end{equation}

\noindent and then the code continues to work as described in \ref{subsec:hammer}.

\subsubsection{Smagorinsky}
\label{subsubsec:sma}

Smagorisky \cite{smago} proposed a model to close the filtered Navier-Stokes equations using the strain-rate tensor of the flow. The model was developed for incompressible flow and hence closing only $\tau_{ij}$, but it can be extended to compressible flow with the closure of $q_j$. The model relies on the eddy viscosity which is computed as:

\begin{equation}
    \nu_t = 2C\Delta^2|\check{S}| 
    \label{eq:sma_nu}
\end{equation}

\noindent where $C=0.0256$ is the Smagorinsky constant, $\Delta$ is the spatial length scale (for our case in computational space), $|\check{S}|=\sqrt{2\check{S}_{ij}\check{S}_{ij}}$ is the norm of the strain-rate tensor, and $\check{S}_{ij}$ is the strain-rate tensor. The strain-rate tensor is computed using the computational space derivatives and hence $\check{S}_{ij}=\frac{1}{2}\left(\frac{\partial \check{u}_i}{\partial \xi_j} + \frac{\partial \check{u}_j}{\partial \xi_i}\right)$. To be consistent with the unit measure also the spatial length scale $\Delta$ needs to be computed in computational space, which can be computed using the Gauss-Legendre quadrature weights $\Delta=\sqrt[3]{w_1w_2w_3}$ where the numbering is the direction. When the mesh is perfectly orthogonal this formulation is the same as if the derivatives are computed in physical space and the spatial length scale is the physical one. 

From the eddy viscosity the SFS stress tensor can be computed as:

\begin{equation}
    \tau_{ij} = -\nu_t(\check{S}_{ij} - \frac{1}{3}\check{S}_{kk}\delta_{ij})
    \label{eq:tau_sma}
\end{equation}

\noindent where $\delta_{ij}$ is the Kronecker delta. The SFS heat flux can be computed as:

\begin{equation}
    q_j = -\frac{\nu_t}{Pr_t}\frac{\partial \check{T}}{\partial \xi_j}
    \label{eq:q_sma}
\end{equation}

\noindent where $Pr_t=0.9$ is the turbulent Prandtl number and it is assumed to be constant in this formulation.

\subsubsection{Dynamic Smagorinsky}
\label{subsubsec:dyn}

Given the limitation of the Smagorinsky model of being bound to the same model for all kinds of turbulence, Moin et al. \cite{dynamicSma} developed a dynamic model which is able to adapt to different kinds of turbulence. The original model was proposed for incompressible flow by Germano et al. \cite{inds} and then extended to compressible flow by Moin et al.. The Dynamic Smagorinsky model is based on the standard Smagorinsky model but instead of assuming $C$ and $Pr_t$ constant, they are computed dynamically with equations--15 and 19 in \cite{dynamicSma}--derived from algebraic identities, which involves a lot of filtering operations. The model was developed to work in fully spectral code and to the best of my knowledge has never been applied to a pseudo-spectral code as H3AMR. Hence, we choose to apply the model to each element. Therefore, each element is treated on its own and it has a specific value of $C$ and $Pr_t$. Because of the small domain covered by the elements it can happen that the constants are negative, therefore the code imposes a lower limit of 0 for $C$ and $0.01$ for $Pr_t$.

\subsubsection{Vreman's model}
\label{subsubsec:vreman}

On the base of the Smagorinsky model Vreman \cite{vreman} developed an Eddy viscosity model specifically for turbulent shear flow. The model estimates $\nu_t$ to close the filtered Navier-Stokes equation but in this case the dissipation is small for transitional and near-wall region, making it particularly suited to be applied in wall bounded or boundary layer flows. The eddy viscosity is computed as:

\begin{equation}
    \nu_t = C_{Vr} \sqrt{\frac{\beta_{11}\beta_{22} - \beta_{12}^2 + \beta_{11}\beta_{33} - \beta_{13}^2 + \beta_{22}\beta_{33} - \beta_{23}^2}{\alpha_{ij}\alpha_{ij}}} 
    \label{eq:vre_nu}
\end{equation}

where

\begin{equation}
    \beta_{ij} = \Delta_m^2\alpha_{mi}\alpha_{mj}
    \label{eq:vre_beta}
\end{equation}

and

\begin{equation}
    \alpha_{ij} = \frac{\partial \check{u}_j}{\partial \xi_i}.
    \label{eq:vre_alpha}
\end{equation}

$C_{Vr}$ is a constant that in this case is set to 0.07 and $\Delta_m$ is the grid spacing along the $m^{th}$ direction, in this to be consistent with our formulation the grid spacing is computed in computatinal space.

\subsubsection{Block Spectral Stresses (BSS)}
\label{subsubsec:BSS}

The Block Spectral Stresses (BSS) method is based on the work of Sousa and Scalo \cite{LSV} and specifically redesigned for pseudo-spectral code. This change of approach is due to the fact that LSV was only validated as a shock capturing method and does not work as LES method because it makes the simulation blow up, even though it is based on a model Quasi Spectral Viscosity (QSV) \cite{QSV}, which is designed for spectral method. The new model computes the estimated cutoff kinetic energy block by block using the gradient of the velocity in each solution point: 

\begin{equation}
    E_N^i(\check{u}_j)=\left| \frac{1}{\gamma_N} \sum_{k=0}^N \left(\ell_j \frac{\partial \check{u}_j(\xi_k^i)}{\partial \xi_i} \right) ^2 L_N(\xi_k^i) w_k \right|
    \label{eq:new_E_BSS}
\end{equation}

\noindent where $\frac{\partial \check{u}_j(\xi_k^i)}{\partial x_i}$ is the computational derivative of $\check{u}_j$ along the $j$ direction computed with the correction polynomials, $L_N(\xi_k^i)$ is the value of the Legendre polynomial at $\xi_k^i$ ($i$ is the direction in which the value is filtered), $w_k$ is the Gauss Legendre quadrature weight, $\ell^j$ is the computational length scale of the solution point (which in the computational space is equal to $w_j$), $\gamma_N=\frac{2}{2N+1}$, and $N$ is the polynomial order of the function used to reconstruct the solution inside the element. BSS method has better shock-capturing capability than LSV and made the SFS quantities more physical, considering that for LSV were not even symmetrical on a 2D vortex in a freestream flow, because the cutoff energy was computed from $\check{u}_j(\xi_k^i)^2$ instead of using the derivative. Indeed the goal of taking only the last mode of the kinetic energy was to remove any freestream interference but considering that the velocity is squared the freestream component is not removed. Instead, the gradient of the velocity removes the freestream interference because it considers the rate of change and not the actual value.

From the estimated cutoff energy, it is possible to compute the subfilter velocity scale $v_i(\check{u}_j)=\sqrt{\frac{N}{2}E_N^i(\check{u}_j)}$, where the factor $N/2$ is the average grid spacing of the element in computational space. From that it is possible to estimate the dissipation needed by the model:

\begin{equation}
    \mathcal{D}_{ij} = v_i(\check{u}_j)\ell_j.
    \label{eq:D_BSS}
\end{equation}

\noindent From the dissipation the SFS quantities can be computed as:

\begin{equation}
    \tau_{ij} = -\frac{1}{2}\left(\mathcal{D}_{ij}\frac{\partial \check{u}_i}{\partial \xi_j} + \mathcal{D}_{ji}\frac{\partial \check{u}_j}{\partial \xi_i}\right)
    \label{eq:tau_BSS}
\end{equation}

\begin{equation}
    q_{j} = -\mathrm{diag}(\mathcal{D})_j\frac{\partial \check{T}}{\partial \xi_j}
    \label{eq:q_BSS}
\end{equation}

\begin{equation}
    \pi_{j} = -\mathrm{diag}(\mathcal{D})_j\frac{\partial \bar{p}}{\partial \xi_j}
    \label{eq:pi_BSS}
\end{equation}

\noindent where $\mathrm{diag}(\mathcal{D})$ is the vector of the diagonal of the matrix and all the indices are not used in Einstein's notation but as the actual index of the vector or matrix. Important consideration is that unlike the Smagorinsky models the diagonal of the strain-rate tensor is not removed to compute the SFS stress tensor because it has an important role when the model is used as shock capturing. Moreover, differently than LSV were the SFS quantities are modulated for this model they are not.

\section{BSS shock capturing capability}
\label{sec:BSS_shock}

Given the fact that BSS is both a shock-capturing and LES method in this section we are going to show its capability as shock-capturing method. The first case use for the investigation is the Sod shock tube (Section \ref{subsec:SOD}) and then the investigation moved to a 2D shock in the shock vortex interaction case (Section \ref{subsec:SVI}).

\subsection{Sod shock tube}
\label{subsec:SOD}

Sod shock tube \cite{sod} is a standard case widely used to test shock-capturing methods \cite{HAGA2019534,TERASHIMA2013484,LSV}. The case simulates a tube with a gas with high pressure and density on the left and lower values on the right separated by a membrane. When the membrane breaks a shock and an expansion wave form propagating respectively to the right and to the left. The problem is described by the Euler equations and as initial conditions:

\begin{equation}
[u_1, p, \rho]=\begin{cases} [0, 1, 1] \ \ \ \mathrm{if}\ x\leq0\\\
[0, 0.1, 0.125]\ \ \ \ \ \ \mathrm{if}\ \ \ x>0
\end{cases}
\label{imp_mass_flux}
\end{equation}

where $u_1$ is the velocity, $p$ the pressure, and $\rho$ the density. The biggest constraint is given by the shock that is difficult to simulate with high order polynomial and hence the need of a shock-capturing method.

\begin{figure}[h!]
    \centering
\includegraphics[width=\textwidth]{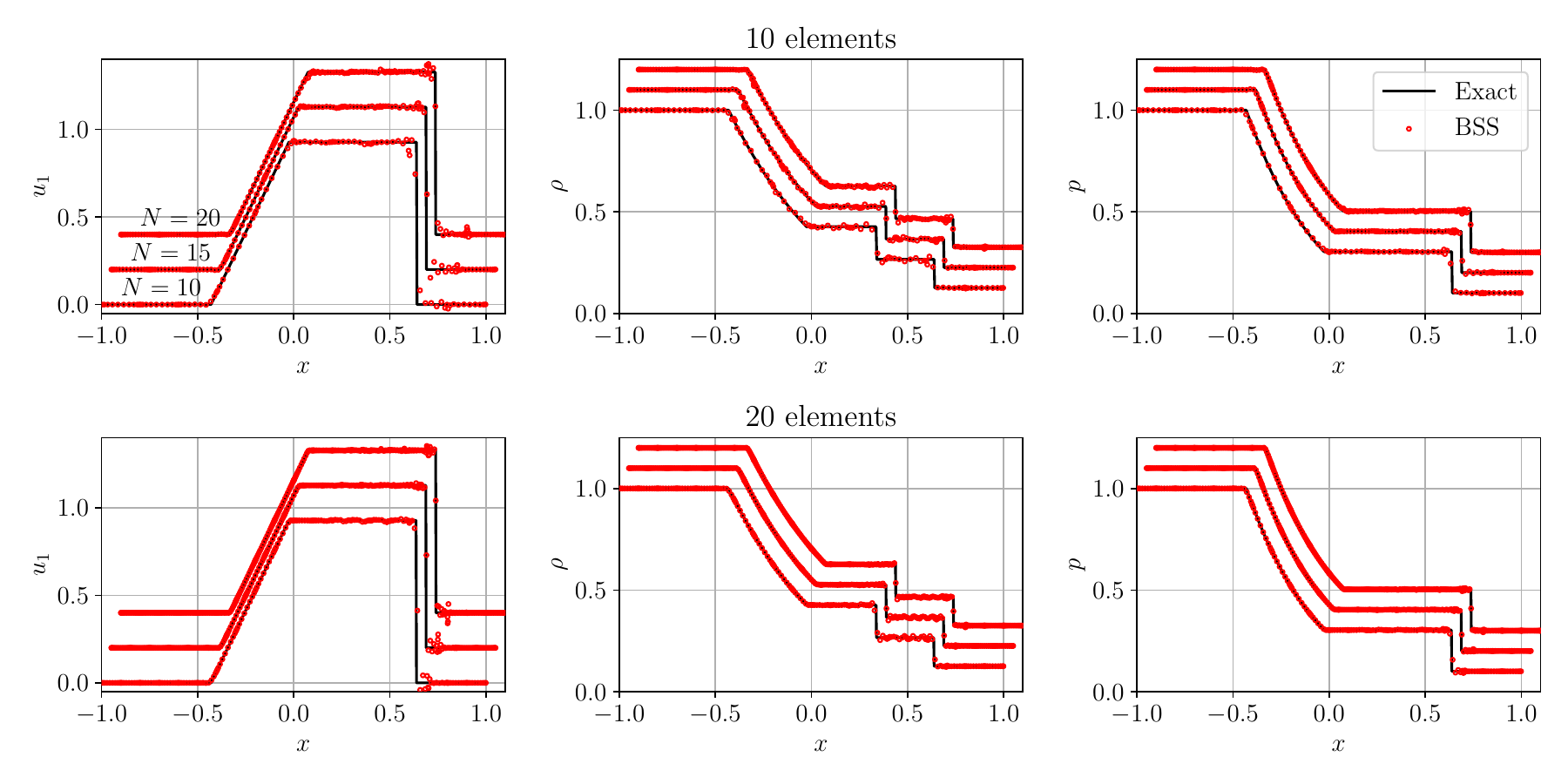}
    \caption{Sod shock tube \cite{sod} velocity $u_1$, density $\rho$ and pressure $p$ at $t=0.264$~s. The first row has a domain of 10 elements and the second of 20 elements. The profiles in all plots from bottom to top used numerical order inside each element 10, 15, and 20 (which correspond to 100, 150, and 200 DOF on the first row and 200, 300, and 400 DOF on the second row).}
    \label{fig:sod_comp}
\end{figure}

Figure \ref{fig:sod_comp} compares the velocity, pressure, and density profiles at $t=0.264$~s of the simulated case with BSS and the analytical solution on different mesh and numerical order $N$ resolution. On the top row the results are obtained using 10 elements and in each plot the numerical order is 10, 15, and 20 from the bottom to the top. Most of the oscillation around the shock are visible in the velocity profile instead the pressure looks mostly smooth. As expected increasing the numerical order allows to have a better resolution of the shock and reduces the oscillation after the shock but on the same time it increases them before the shock. However this phenomenon is not depended on the shock but on the fact that shock is closer to the left face of the element, indeed also on the right face it is possible to see some oscillation. On the lowest row of Figure \ref{fig:sod_comp} are presented the results obtained on a mesh with 20 elements. As before the numerical order used is 10, 15, and 20 from the bottom to the top. The most visible outcome is the fact that the oscillation are reduced when compare to the same numerical order on a lower elements mesh and also allows a better resolution of the shock. Therefore, it is possible to conclude that increasing the number of degrees of freedom (DOF) reduces the oscillation cased by the shock, but to have a better resolution of the shock it is better to increase the number of elements and not only the order.

\begin{figure}[h!]
    \centering
\includegraphics[width=\textwidth]{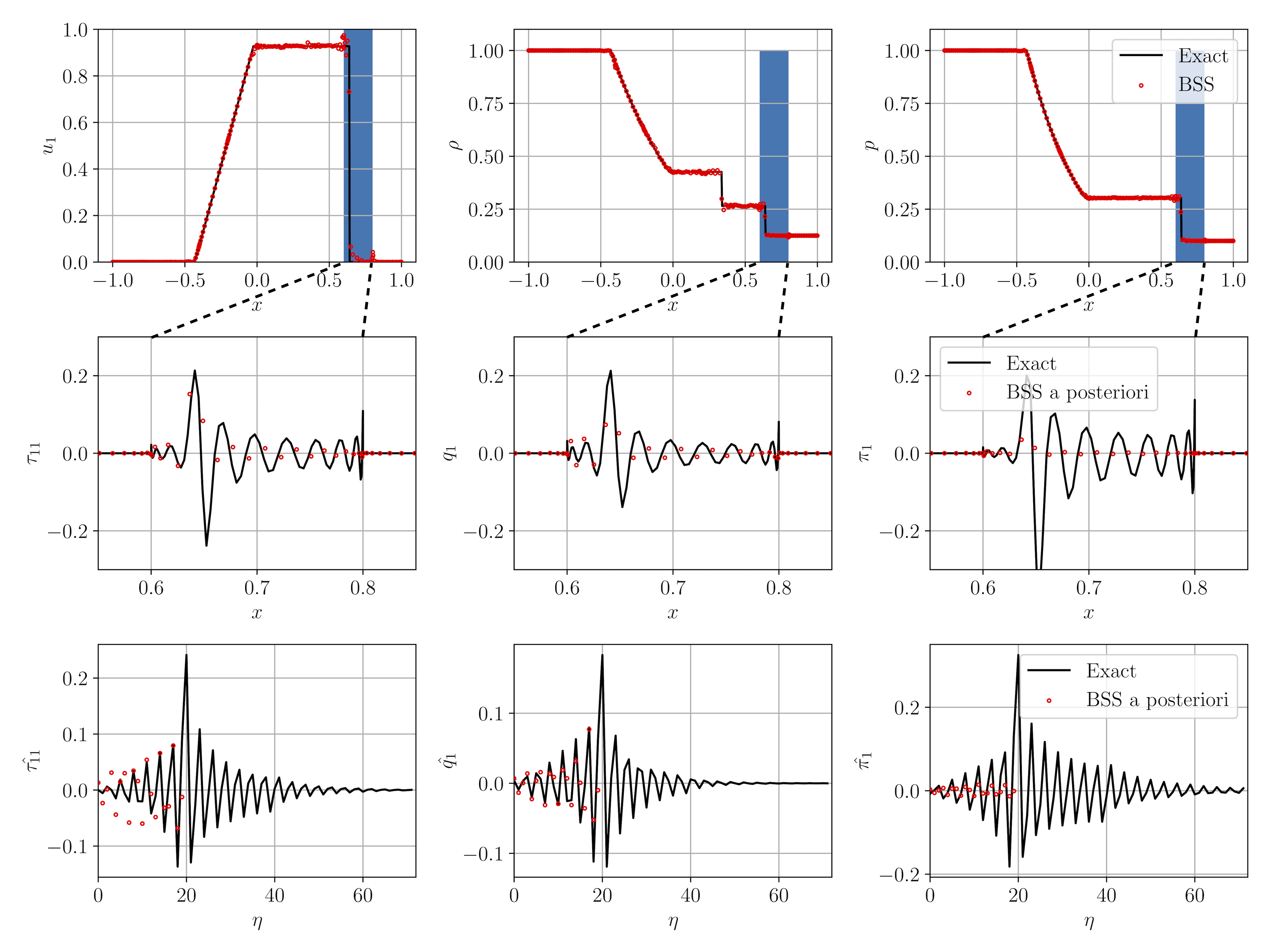}
    \caption{Sod shock tube \cite{sod} velocity $u_1$, density $\rho$, pressure $p$, SFS shear stress $\tau_{11}$, SFS heat flux $q_1$, and SFS pressure-work $\pi_1$ at $t=0.264$~s, for a 10 elements mesh with numerical order 20. The SFS are computed using equations \ref{eq:tau_BSS}-\ref{eq:pi_BSS}. The last row of plots presents the spectral content--computed using the Legendre coefficients--of the three SFS quantities computed in the element with the shock.}
    \label{fig:sod_SFS}
\end{figure}

Figure \ref{fig:sod_SFS} shows the velocity, pressure, density, SFS shear stress, SFS heat flux, and SFS pressure-work for the 10 elements and numerical order 20 case. The exact SFS quantities are computed using equations \ref{eq:tau_BSS}-\ref{eq:pi_BSS} and the BSS one are the a posteriori results of a simulation. As expected considering that the shock has an infinite spectra the SFS quantities have a lot of oscillations and the modeled quantities are able to precisely match the analytical quantities. The last row of plots presents the spectral content computed using the Legendre coefficients and then purely real of the SFS shear stress, SFS heat flux, and SFS pressure-work in the element with shock. For all cases, the a posteriori spectra match the behavior of the exact solution spectra, with the oscillation of sign and similar magnitude. There are some differences in the spectra of the SFS shear stress but it is possible to see that this difference is due to the discrete nature of the investigation, indeed if the results were continuous the exact solution spectra would oscillate further.

\subsection{Shock vortex interaction}
\label{subsec:SVI}

The shock vortex interaction is a case that investigates the interaction between a vortex without circulation and a standing shock. For this investigation the domain is $\left[0,2L\right] \times \left[0,L\right]$, the freestream velocity before the shock is $V_0=1.5\sqrt{\gamma p_0/\rho_0}$, the shock location is at $x_s/L=1/2$, and the vortex center at $x_{cv}/L=1/4$ and $y_{cv}/L=1/2$. The vortex is initialized with the tangential velocity that scales linearly respect to the radius in the inner part and a combination of linear and inverse in the external part. The inner part radius is $a/L=0.075$ and the outer part $b/L=0.175$ the velocity of the vortex can be written as $\mathbf{u}=u_\theta(r)\hat{\mathbf{e}}_\theta+V_0\hat{\mathbf{e}}_x$ where:

\begin{equation}
 \frac{u_{\theta}(r)}{u_{\theta}(a)} = 
 \begin{cases}
 \frac{r}{a} ,& \text{if } r \leq  a\\
\frac{\eta}{2}\left( \frac{r}{b} -  \frac{b}{r}\right)  ,& \text{if } a < r  \leq b\\
 0,& \text{otherwise,}
 \end{cases}
\end{equation}

where $\eta = 2(b/a)/[1 - (b/a)^2]$ and the maximum tangential velocity is $u_{\theta}(a) = 0.9V_0$. The pressure is assumed for ideal gas and isentropic compression and it is initial so that the gradient balance the centrifugal forces:

\begin{equation}
\frac{\partial P}{\partial r} = \rho \frac{{u}_{\theta}^2(r)}{r}, \quad P = \rho R T, \quad \frac{P}{P_0} = \left(\frac{\rho}{\rho_0}\right)^{\gamma}.
\end{equation}

\begin{figure}[h!]
    \centering
\includegraphics[width=\textwidth]{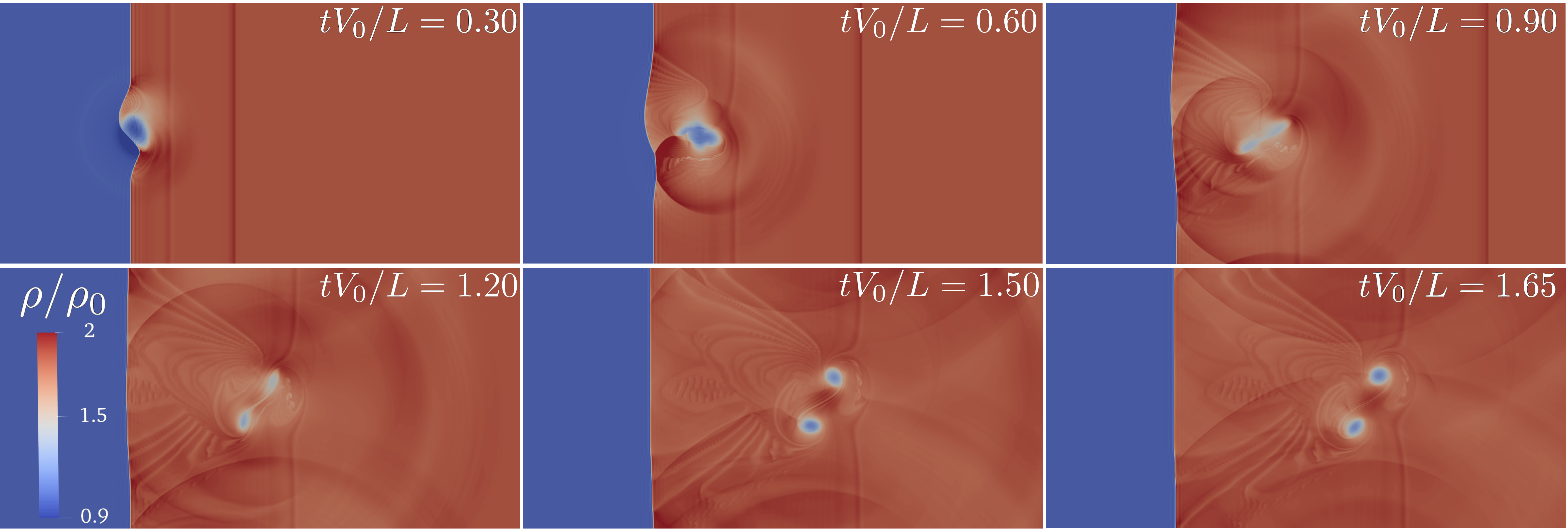}
    \caption{Shock-vortex interaction evolution using numerical order 4 and a mesh 256x128 blocks (524,768 DOF).}
    \label{fig:sv_evolution}
\end{figure}

Other works \cite{sv_ellzey,sv_rault,TONICELLO2020104357} run simulations on the configuration and reported that the interaction of the vortex with shock leads to the creation of two smaller vortices with the upper one in a forward position respect to the lower. Figure \ref{fig:sv_evolution} shows the evolution of the vortex passing through the shock deforming the shock into an S-shape. The vortex is splitted in two smaller vortices that continue to travel to the right and as described in previous works the upper one is in a forward position. This test is important because it tests the shock-capturing capability also on an oblique shock and the same time it does not dissipate the vortex.

\begin{figure}[h!]
    \centering
\includegraphics[width=\textwidth]{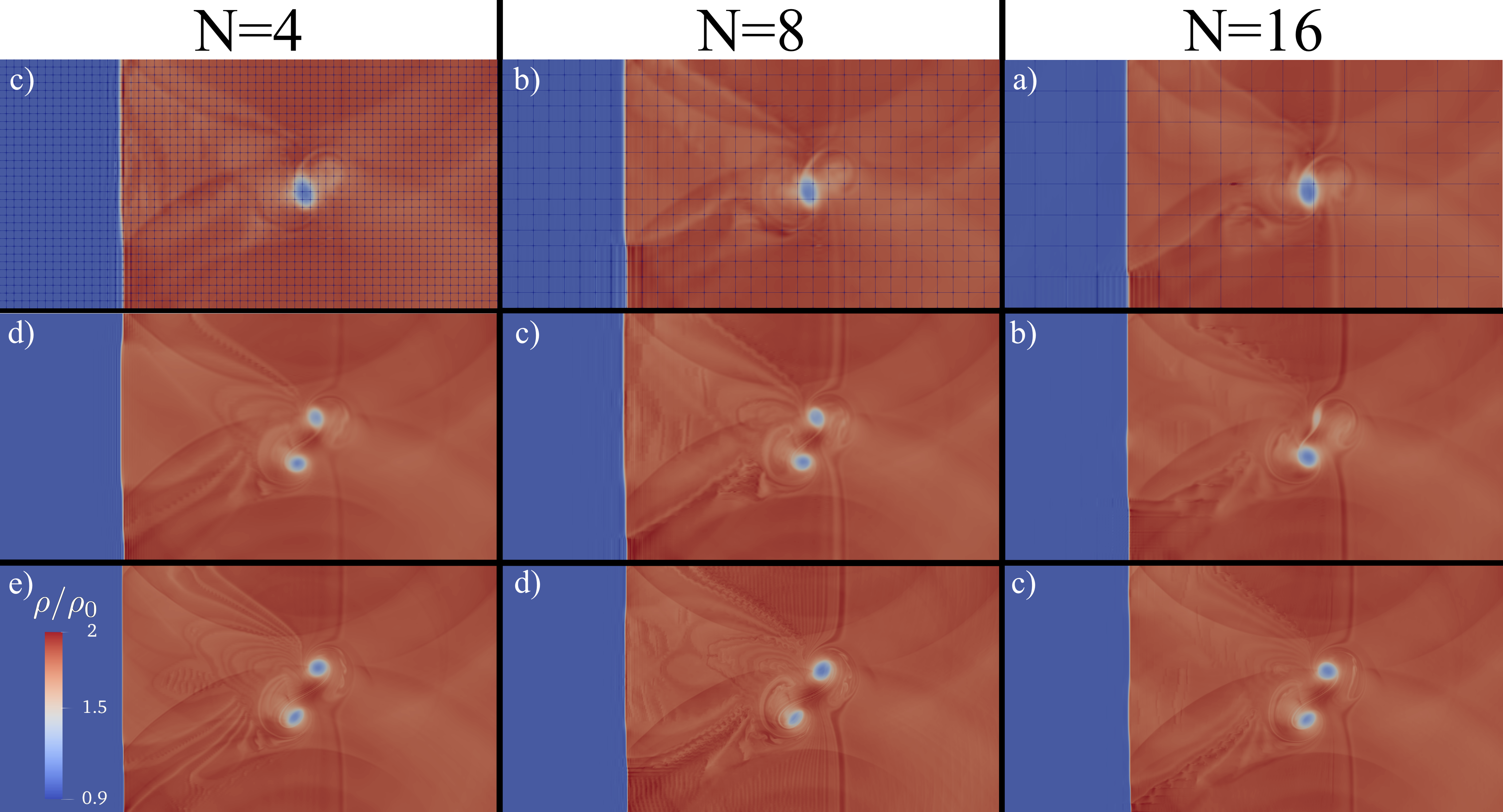}
    \caption{Shock-vortex interaction at $tV_0/L=1.65$ with different numerical order and mesh resolution. Each row have different degrees of freedom in the simulation, respectively 32768, 131072, and 524288. The meshes have in x and y: a) 16x8, b) 32x16, c) 64x32, d) 128x64, and e) 256x128.}
    \label{fig:sv_comp}
\end{figure}

Figure \ref{fig:sv_comp} tabulates a comparison of of different numerical order (in the raws) and the their refinementment (in the columns). As previously seen in the Sod shock tube investigation when the number of degrees of freedom is the same, the simulation with the lower numerical order has the best result. This is reported in the 131,072 DOF case where with numerical order 16 the upper vortex is completely dissipate, instead with order 4 is fully resolved. With 524,288 DOF the resolution of all numerical order reach convergence and they show similar results. Therefore, it is possible to conclude that the method works better when using more elements, this was shown in the case of a shock when looking at the Sod shock tube simulation and now it is fatherly validate also in the case of vortices.

\section{LES comparison}
\label{sec:LES_comp}

\subsection{Taylor-Green Vortex (TGV)}
\label{subsec:TGV}

The Taylor-Green Vortex (TGV) is a 3D flow in a cube in which the flow tends to a state of homogeneous turbulence. Even though there exists an analytical solution for the incompressible case, considering that in this investigation the code used simulates compressible flow the results are compared with a DNS of the flow. The flow starts at $t=0$~s and ends at $t=200$~s in a cubed box $[-\pi L, \pi L]^3$ with the following initial conditions:

\begin{equation}
    \rho(\mathbf{x},0)/\rho_0 = 1,
\end{equation}

\begin{equation}
    \frac{u_1(\mathbf{x},0)}{V_0} = \mathrm{sin}\left(\frac{x_1}{L}\right)\mathrm{cos}\left(\frac{x_2}{L}\right)\mathrm{cos}\left(\frac{x_3}{L}\right),
\end{equation}

\begin{equation}
    \frac{u_2(\mathbf{x},0)}{V_0} = -\mathrm{cos}\left(\frac{x_1}{L}\right)\mathrm{sin}\left(\frac{x_2}{L}\right)\mathrm{cos}\left(\frac{x_3}{L}\right),
\end{equation}

\begin{equation}
    \frac{u_3(\mathbf{x},0)}{V_0} = 0,
\end{equation}

\begin{equation}
    \frac{p(\mathbf{x},0)}{\rho_0 V^2_0} = \frac{p_0}{\rho_0 V^2_0} + \frac{1}{16} \left[\mathrm{cos}\left(\frac{2x_1}{L}\right) + \mathrm{cos}\left(\frac{2x_2}{L}\right)\right]\left[\mathrm{cos}\left(\frac{2x_3}{L}\right)+2\right]
\end{equation}

\noindent where $\rho_0=1$, $V_0=0.1$, $p_0=1/\gamma$, and $L=1$ are the non-dimensionalized density, velocity, pressure, and length. The Mach number of the flow is $M_0=\frac{V_0}{\sqrt{\gamma p_0/\rho_0}}=0.1$, which can be considered incompressible but the solution completely diverges from the analytical incompressible solution. The Reynolds number is $Re=\frac{\rho_0 V_0 L}{\mu_0}=5000$ where the viscosity is considered to be constant at $\mu_0=2\cdot10^{-5}$. All the simulations use a third-order Runge-Kutta time advancement method with $CFL=0.1$. Grid convergence is achieved a $32^3$ elements mesh running with numerical order 9. 

\begin{figure}[h!]
    \centering
\includegraphics[width=\textwidth]{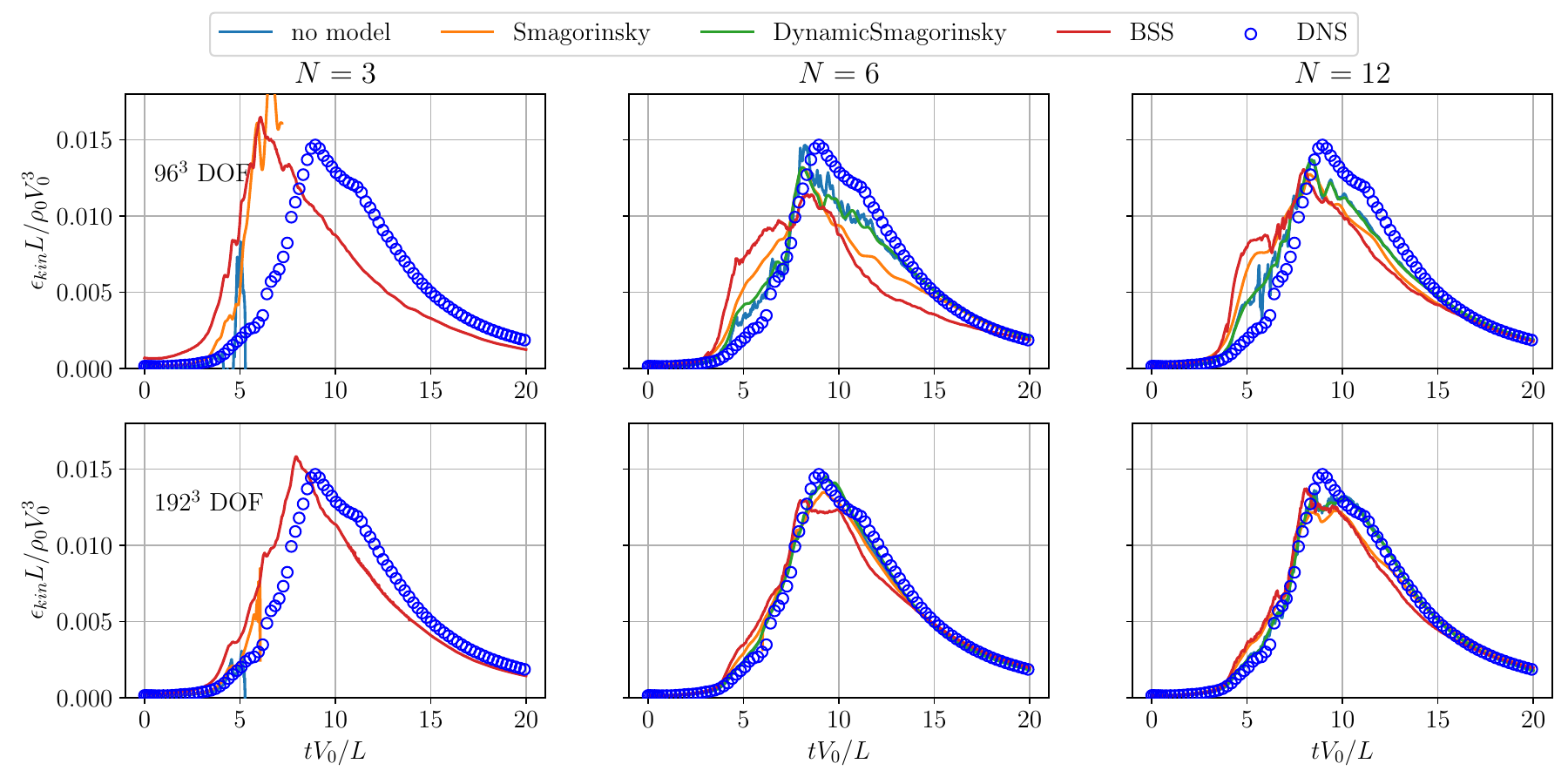}
    \caption{Normalized time and dissipation $\epsilon_{kin}$ with different degrees of freedom (DOF) and numerical order. From left to right the meshes use for the $96^3$ DOF case have respectively $32^3$, $16^3$, and $8^3$ elements. For the $192^3$ DOF respectively $64^3$, $32^3$, and $16^3$ elements.}
    \label{fig:comp_TGV}
\end{figure}

Similarly to the shock-vortex interaction case, figure \ref{fig:comp_TGV} reports how different methods behaves with different degrees of freedom (DOF) and numerical orders. On the y-axis there is the normalized dissipation computed as $\epsilon_{kin}=-\frac{\partial E_{kin}}{\partial t}$ and on the x-axis the normalized time. When using numerical order 3 only BSS is able to resolve the turbulence instead the order method starts to oscillate and then blow up. Differently from spectral method where the blow up is caused by the dissipation becoming positive, in block spectral code the blow up is caused by oscillations in the single block which in some cases can be seen in the general statics as the dissipation. When running with $96^3$ DOF the BSS is significantly more dissipative than the other method when the turbulence starts to break down and consequently the peak dissipation is smaller than the other methods. In the $192^3$ DOF case BSS is the only method able to capture the plateau after the peak instead the other methods have a single peak or another peak after the first one where the plateau should be. The dynamic Smagorinsky method seems a filtered version of the case without any model, this is caused by the fact that the model is applied by block by block so the effect on the simulation is small but strong enough to reduce the oscillation of not having a model.

\begin{figure}[h!]
    \centering
\includegraphics[width=0.8\textwidth]{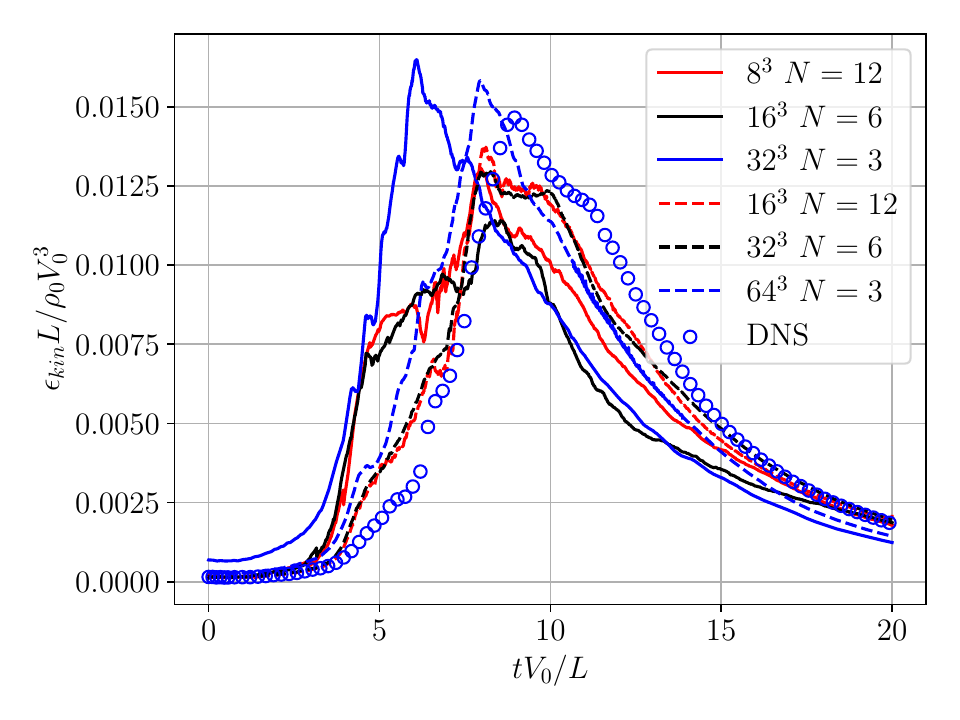}
    \caption{Dissipation convergence of different mesh and numerical resolution when using the BSS method. Solid lines indicate cases with $96^3$ DOF and dashed lines $192^3$ DOF. Red, black, and blue color indicate simulation respectively with numerical order 12, 6, and 3.}
    \label{fig:BSS_TGV}
\end{figure}

Figure \ref{fig:BSS_TGV} plots the dissipation of the different mesh and numerical order so to highly how the method perform with different setup. All the cases with $96^3$ DOF have very different dissipation indeed in the $32^3$ order 3 the peak dissipation happens sooner and is stronger than what it should be. The order 12 case is the one with the best performance due to the fact that in this case the model is able to be more active and is able to better capture the turbulence. The cases with $192^3$ DOF are closer to the DNS resolution and for this reason the mesh or the order do have a big different has it has been seen in the shock-vortex interaction case. The biggest difference is the peak that is stronger for the order 12 simulation thanks to a bigger spectra available to the model.

\begin{table}[!h]
\caption{Time, number of iterations, and the second per iteration all normalized respect to the no model case for TGV simulation with no model, Smagorinsky, Dynamic Smagorinsky, and Legendre Spectral Viscosity (BSS) methods. The results are obtained on a $16^3$ elements mesh and numerical order 6.}
\centering
\begin{tabular}{l|ccc}
\hline\noalign{\smallskip}
 & Time & \char"0023 \ iterations & Time per iteration \\
\noalign{\smallskip}\hline\noalign{\smallskip}
No model & 1.000 & 1.000 & 1.000 \\
Smagorinsky & 1.009 & 0.987 & 1.024 \\
Dynamic Smagorinsky & 0.940 & 0.801 & 1.173 \\
BSS & 1.067 & 0.990 & 1.078 \\
\noalign{\smallskip}\hline
\end{tabular}
\label{tab:time}  
\end{table}

Table \ref{tab:time} reports the computational time, the number of iterations, and the time per iteration by all LES models and without any model, on a $16^3$ elements mesh with numerical order 6. As expected the Dynamic Smagorisky model is the slowest due to the big number of filtered quantities that it needs to estimate the coefficients. The standard Smagorisky is the fastest considering that it only needs to multiply a constant with the gradients of the velocity and the temperature. BSS is placed in the middle of the two Smagorisky models being 6\% slower than the standard one. From the number of iterations it is possible to conclude that all models lead to a lower time constraint induced by the SFS stresses because all of them require fewer iterations than without a model.

\subsection{Supersonic channel flow}
\label{subsec:channel}

\begin{figure*}[h!]
    \centering
    \vspace*{.5in}
\includegraphics[width=0.98\textwidth]{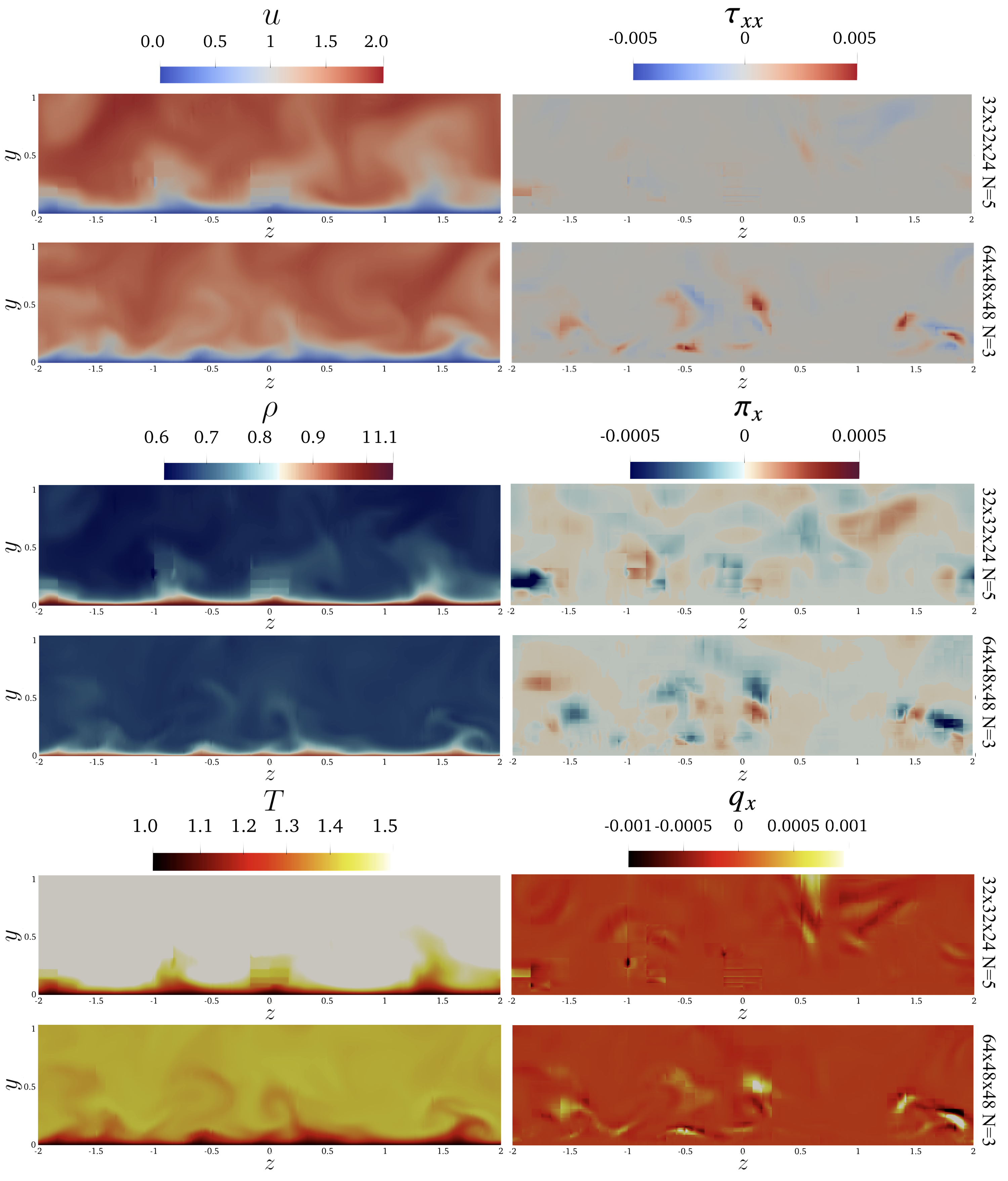}
    \caption{y-z section of channel flow mach 1.5 and no-slip isothermal boundary condition at the wall. On the left column the velocity, density, and temperature contours and on the right one the SFS shear stress, pressure-strain, and heat flux. The first row of each quantity is on a 32x32x24 mesh and numerical order 5. The second is on a 64x64x48 mesh with numerical order 3.}
    \label{fig:channel_cut}
\end{figure*}

A compressible turbulent channel flow has been used to check BSS performance in wall bounded flow, specifically in a supersonic case Mach 1.5 and an hypersonic case Mach 6.0. The flow condition match the one used by Chen and Scalo \cite{chen_scalo_2021}, all the flow parameters are normalized using the channel's half-width ($\delta$), the speed of sound at the wall ($a_w$), the wall temperature ($T_w$), and the bulk density $\rho_w$. The results are plotted using the Trettel-Larsson transformation \cite{trettel2016mean}, where the mean velocity is transformed as:

\begin{equation}
    U^+_{TL} = \int_0^{\check{u}/u_{\tau}}\sqrt{\frac{\overline{\rho}}{\rho_w}}\left(1+\frac{1}{2}\frac{1}{\overline{\rho}}\frac{\partial\overline{\rho}}{\partial y}y - \frac{1}{\overline{\mu}}\frac{\partial \overline{\mu}}{\partial y} y\right)d\left(\frac{\check{u}}{u_{\tau}}\right)
    \label{eq:U_TL}
\end{equation}

where $u_{\tau}=\sqrt{\overline{\tau_w}/\overline{\rho_w}}$, plotted versus:

\begin{equation}
    y^*=\frac{\overline{\rho}(y)u^*_{\tau}}{\overline{\mu}(y)}y
    \label{eq:y_star}
\end{equation}

where $u_{\tau}=\sqrt{\overline{\tau_w}/\overline{\rho}(y)}$. The reference log-law profile is computed as:

\begin{equation}
    U^+_{TL}=\frac{1}{\kappa}\mathrm{ln}(y^*)+C
    \label{eq:log-law}
\end{equation}

where $\kappa=0.41$ and $C=5.5$.

\begin{figure}[h!]
    \centering
\includegraphics[width=1.0\textwidth]{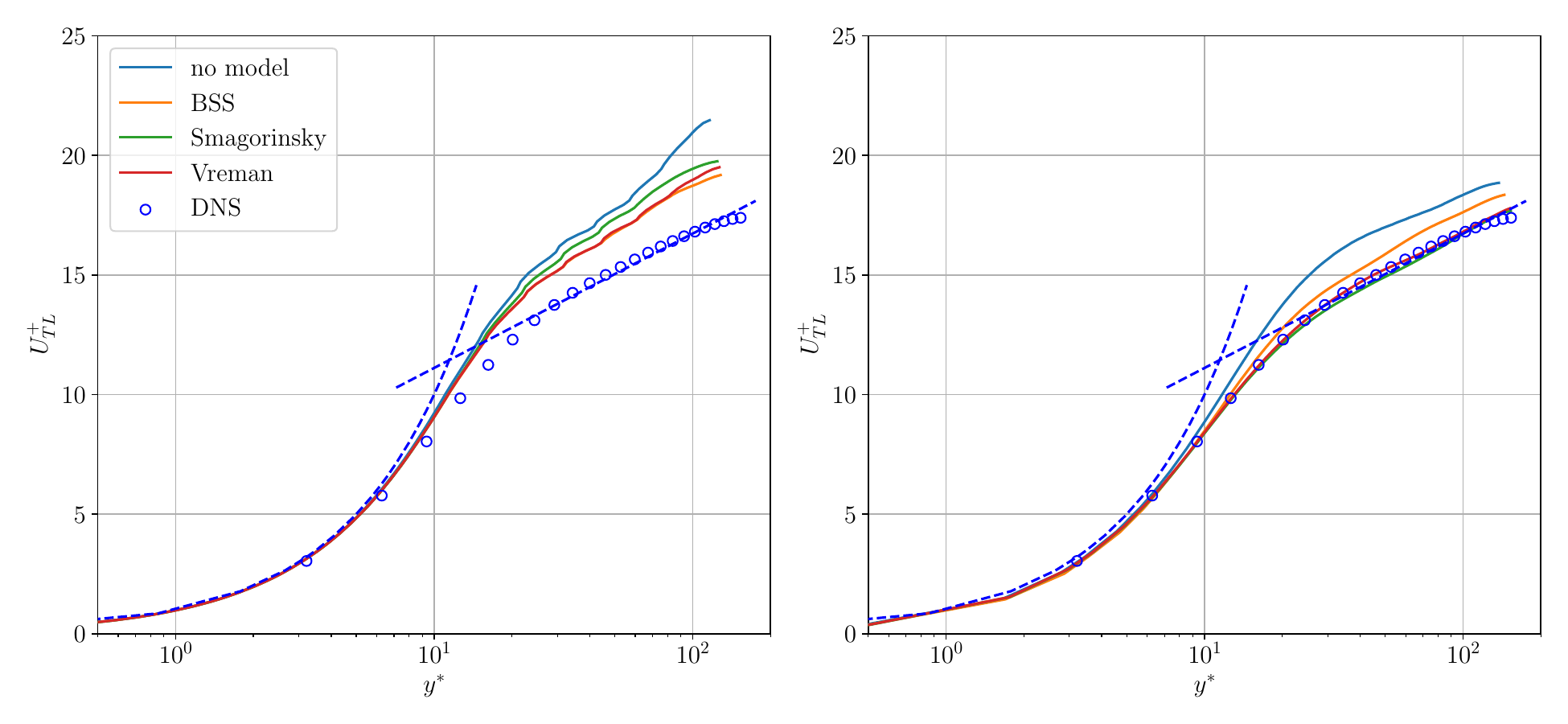}
    \caption{TL-transformed mean velocity profile for Mach 1.5 compressible channel flow. On the left using a 32x32x24 elements mesh numerical order 5 and on the right a 64x48x48 mesh numerical order 3.}
    \label{fig:channel_1.5}
\end{figure}

Figure \ref{fig:channel_cut} shows on the left column the velocity, density, and temperature contours and on the right one the SFS shear stress, pressure-strain, and heat flux for a 64x48x48 mesh order 3 and 32x32x24 mesh order 5. It is possible to see that the finer mesh is able to better resolve the boundary layer and the flow looks more continuous, instead in the coarse mesh the boundary layer is thicker and sometimes the flow looks discontinuous cell by cell but this problem is mainly due to Paraview interpolation and partially to the block spectral nature of the solver. The SFS quantities are not as continuous as the flow also for the finer mesh case in this instance the effect is because BSS computes these quantities block by block and does not enforce continuity. It is interesting to notice that the SFS quantities have a stronger magnitude on the finer mesh case where the numerical order is lower and they are weaker with high numerical order. This result agrees with the fact that a high numerical order is able to better resolve the dynamic properties of the flow.

Figure \ref{fig:channel_1.5} presents more quantitative results of the same case with the plots of the TL-transformed mean velocity. On the left has been used a 32x32x24 elements mesh with numerical order 5, it is possible to see that the BSS model has the best performance together with Vreman's model indeed the Smagorinsky model is not able to reduce the velocity as much as BSS. On the right has been used a finer mesh with 64x48x48 elements and numerical order 3. In this case Smagorinsky and Vreman model precisely match the DNS results instead BSS do not fully resolve the turbulence in the log-law layer. The fact that BSS is not able to be as good as the other models when using a finer mesh with lower numerical order can be also explained by the SFS terms shown in Figure \ref{fig:channel_cut} where the SFS quantities are stronger when using lower numerical order. Indeed BSS is more related to the numerical order then the other two methods because to compute the SFS energy dissipation it needs the energy spectrum. Instead Smagorinsky and Vreman models depend on the numerical order only by the grid spacing. 

\begin{figure}[h!]
    \centering
\includegraphics[width=0.6\textwidth]{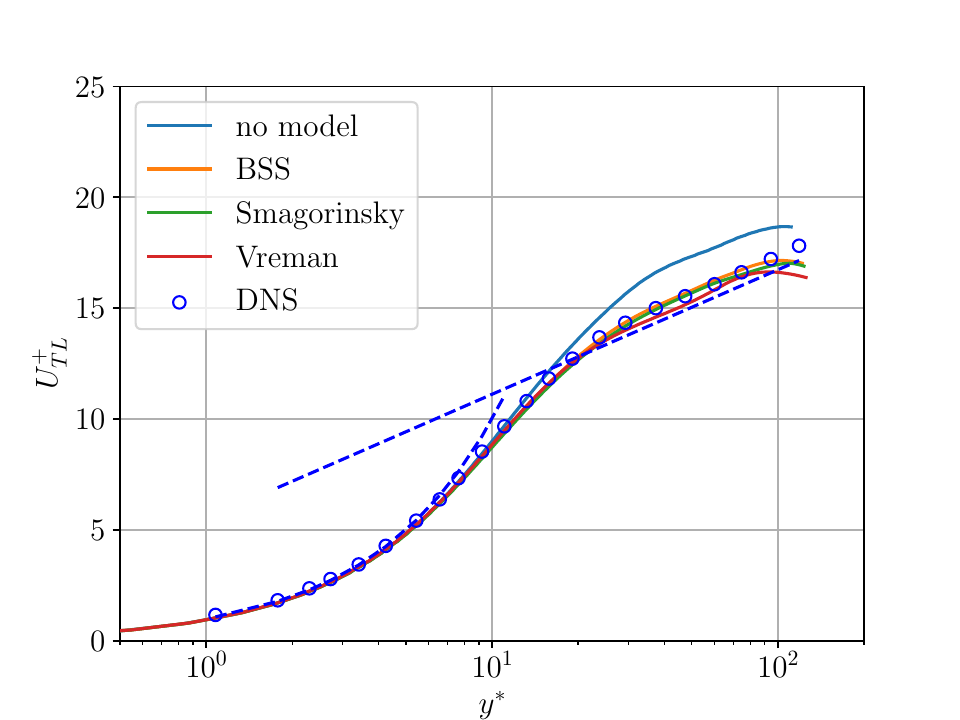}
    \caption{TL-transformed mean velocity profile for Mach 6.0 compressible channel flow. On the left using a 96x75x48 elements mesh numerical order 3.}
    \label{fig:channel_6}
\end{figure}

Figure \ref{fig:channel_6} plots the results for the Mach 6.0 hypersonic case. The mesh used for the simulation has 96x75x48 elements and uses numerical order 3. For this case the Smagorinsky and BSS model leads to a similar solution that match the DNS results, instead Vreman's model predictis a lower velocity closer to the log-law profile. In both the Mach 1.5 and 6.0 the Dynamic Smagorinsky did not work probably because the averaging is performed element by element and not by layer as suggested by the authors for this kind of simulation. Considering that the averaging have to variate case by case it would be difficult both to tune and to code in a CFD solver and hence we can conclude that Smagorisky, Vreman and BSS are versatile enough to work with different flow condition in a block spectral code.

\section{Conclusion}
\label{sec:conclusion}

In this paper, we presented a new shock-capturing and LES model for high-order finite volume methods called Block Spectral Stresses (BSS) estimation. The method relies on the last mode of the velocity gradient inside each element to estimate the subfilter scale stresses, heat-flux, and pressure-work to close the filtered Navier-Stokes equations. The main idea is that both turbulence and shock formation have a cascade of energy from large to small scale and hence a closure for the turbulence model should work also for shock-capturing. Specifically, we noticed that in our model the subfilter scale pressure-work, which is rarely modeled in LES closure, is really important to capture the shock.

BSS has been implemented in a block spectral unstructured code called H3AMR and validated on simulations with shocks and turbulence. In the Sod shock tube case, the model allows to simulate up to numerical order 25, we were not able to reach higher order because of the limitation but the model can potentially allow higher orders. From this investigation, we also notice how the shock is better predicted when the mesh is refined and not as much when the numerical order is increased. Another shock investigation was on the shock-vortex interaction case, where the model demonstrated to be able to capture the shock without altering the shape of the vortex. Also in this case, the mesh refinement leads to faster convergence than when increasing the numerical order.

For the LES closure, BSS has been compared to the Smagorinsky, dynamic Smagorinsky, and Vreman models, with the second changed to work in a block spectral code. The Taylor-Green vortex was used to test the model in a homogeneous isotropic simulation. In the case of numerical order 3 BSS is the only model able to estimate the small scale turbulence and avoid the simulation blow-up. For numerical order 6 and 12 and coarse mesh BSS is more dissipative than the other methods but on finer meshes, it is able to reach similar results to the other models. Another turbulence simulation was performed for supersonic and hypersonic channel flow; in this case, dynamic Smagorinsky does not work because of the averaging chosen for the block spectral implementation, instead BSS, the standard Smagorinsky, and Vreman were able to simulate the flow. Differently, from TGV in the supersonic case, BSS achieved better performance at predicting the velocity profile when using a coarser mesh with higher numerical order because of a better prediction of the SFS terms. For the hypersonic case, Smagorinsky is slightly closer to the DNS result than BSS but the difference is minimal, instead Vreman's model estimates a lower velocity closer to the log-law profile instead of the DNS.

Therefore, we can conclude that even though BSS is not the best model in every kind of flow is able to work with every flow without big changes in the code, as in the case of dynamic Smagorinsky which requires different averaging for different kinds of flows. Currently, the model has been used with flux reconstruction numerics a Legendre base but we expect that the model can work also with different numerics and base on high-order finite volume. This last point requires further investigation.

\bibliographystyle{unsrt}
\bibliography{biblio}

\end{document}